\documentclass[12pt]{iopart}
\usepackage{iopams} 
\usepackage[normalem]{ulem}
\usepackage{soul,xcolor}
\setstcolor{red}
\usepackage{graphicx}  
\begin{document}

\title[Localized exciton-polariton modes in dye-doped nanospheres]{Localized exciton-polariton modes in dye-doped nanospheres: a quantum approach}

\author{Martin J Gentile, Simon A R Horsley, William L Barnes}
\address{School of Physics and Astronomy, University of Exeter, Exeter EX4~4QL, UK}
\ead{\mailto{m.j.gentile@exeter.ac.uk}, \mailto{s.horsley@exeter.ac.uk}, \mailto{w.l.barnes@exeter.ac.uk}}

\begin{abstract}
We model a dye-doped polymeric nanosphere as an ensemble of quantum emitters and use it to investigate the localized exciton-polaritons supported by such a nanosphere. By determining the time evolution of the density matrix of the collective system, we explore how an incident laser field may cause transient optical field enhancement close to the surface of such nanoparticles. Our results provide further evidence that excitonic materials can be used to good effect in nanophotonics.
\end{abstract}

\pacs{7135Gg, 4250Ct, 7135Cc, 7722Ch, 4250Md}

\vspace{2pc}

\submitto{\JOPT}
%
\maketitle
%
%

\vspace{10pt}

\section{Introduction}
Plasmonic nanoparticles exhibit optical field enhancement when localized surface plasmon polariton (SPP) resonances are excited~\cite{LRandE,Vollmer}. The strength of the enhancement depends sensitively on the nanoparticle's environment and geometry~\cite{Kelly_JPCB_2003_107_668}. The enhancement is a vital part of phenomena such as surface-enhanced Raman scattering~\cite{Stiles_AnnRevAnalChem_2008}, and finds application in areas such as biosensing~\cite{Willets_AnnuRevPhysChem_2007_58_267}, monitoring lipid membranes~\cite{Taylor_SciReps_2014_4_5940}, modifying molecular fluorescence~\cite{Kitson_PRB_1995_52_11441} and materials characterization~\cite{Isaac_APL_2008_93_241115}. Localized SPP resonances occur because of the way the free conduction electrons in metal particles respond to light. For many metals at optical frequencies their response is such that the permittivity is negative - a critical requirement if the nanoparticle is to support a plasmon mode.\\

However, metals are not the only materials to exhibit negative permittivity; materials doped with excitonic organic dye molecules are of interest for photonics~\cite{Saikin_Nanophotonics_2013_2_21} and may also possess negative permittivity over a small frequency range~\cite{Philpott_MolCrystLiqCryst_1979_50_139}. Interest in such materials as a means to support surface exciton-polariton (SEP) resonances has recently been rekindled~\cite{Gu_APL_2013_103_021104,Gentile_NL_2014_14_2339,Triolo_arxiv1503_07499}. An example of this class of material is a polymer doped with dye molecules. In a previous work we showed, through experiment and with the aid of a classical model, that polyvinyl alcohol (PVA) doped with TDBC molecules (\textit{5,6-dichloro-2-[[5,6-dichloro-1-ethyl-3-(4-sulphobutyl)-benzimidazol-2-ylidene]-propenyl]-1-ethyl-3-(4-sulphobutyl)-benzimidazolium hydroxide}, sodium salt, inner salt) may support localized surface exciton-polariton modes. We extracted the complex permittivity $\varepsilon(\omega)$ of this material from reflectance and transmittance measurements of thin films using a Fresnel approach~\cite{AandB}. TDBC was chosen because of its tendency to form J-aggregates: this leads to a narrowing of the optical resonance, making them interesting for strong coupling~\cite{Lidzey_Science_2000_288_1620,Dintinger_PRB_2005_71_035424, Torma_RepProgPhys_2015_78_013901}. More importantly in the present context, at sufficiently high concentrations materials doped with such molecules exhibit a negative permittivity; it is this negative permittivity that enables these materials to support localized resonances. In our previous work~\cite{Gentile_NL_2014_14_2339} a two-oscillator Lorentz model~\cite{Fox_OPS,lebedev} was used to calculate the electric field enhancement and field confinement around the nanoparticle supporting the resonance by use of Mie theory~\cite{Mie,B+H}. The field enhancement and confinement we predicted compared favorably with respect to gold nanospheres, albeit over a much narrower spectral range. Here we extend that earlier work by going beyond a simple classical Lorentz oscillator model. In doing so, we are able to explore new transient phenomena and develop a richer microscopic physical picture for the system.\\

In what follows we first outline the elements and assumptions of the quantum model we have used. We then compute the relative permittivity, $\varepsilon(\omega)$, using this model and compare our results with those obtained from a classical model. We next use our model to investigate the steady-state response of nanospheres of possessing this relative permittivity, with a focus on the nature of the localized surface exciton-polariton (LSEP) mode. The response of the same particle to a suddenly turned on applied optical field is then explored, and the occurrence of transient LSEP modes discussed.\\

\section{Theory}\label{sec:theory}
The key difference between the work we report here and previous work based on a bulk, macroscopic approach~\cite{Gentile_NL_2014_14_2339} is that here we develop an effective medium description from a quantum model of the relative permittivity $\varepsilon=1+\chi$, where $\chi$ is the susceptibility. To do so we assume that the molecules in our material can be represented as an ensemble of two-level quantum systems. For a general material, an applied electric field $\bi{E}$ induces a polarization in the material $\bi{P}\propto{\langle\bi{d}\rangle}$, where $\langle\bi{d}\rangle$ is the average dipole moment of each molecule (quantum system). Assuming linearity, $\bi{P}$ is a linear function~\cite{Mandel_Wolf} of $\bi{E}$, given by:

\begin{equation}
	\bi{P} = \frac{1}{2}\varepsilon_0\bi{E}(\chi e^{-i\omega t}+\chi^*e^{i\omega t}) = N\langle\bi{d}\rangle
\label{eq:Polarisation1}
\end{equation}

\noindent where $N$ is the number density of quantum emitters, and $\bi{E}$ and $\bi{d}$ are generally time and frequency dependent. To find $\chi$ and hence $\varepsilon$ we need to find $\langle\bi{d}\rangle$. If we adopt a quantum picture, then $\langle\bi{d}\rangle$ becomes the expectation value of the dipole moment and can be computed from the trace of $\rho\bi{d}$, where $\rho$ is the density matrix of the system and $\bi{d}$ is the transition dipole matrix. The density operator $\hat{\rho}$ is defined as $\hat{\rho}=\sum_k p_k|k\rangle\langle k|$, where $p_k$ are the relative probabilities of finding a system element in state $|k\rangle$. In order to find $\rho$, a Hamiltonian that describes the system must be determined. In general, the Hamiltonian for an open quantum system can be expressed as~\cite{Skinner_JPhysChem_1986,Abramavicius_Wurfel_2011},

\begin{equation}
	\hat{H}=\hat{H}_0+\hat{H}_B+\hat{H}_I,
\end{equation}

\noindent where $\hat{H}_0$ is the Hamiltonian of the isolated system, $\hat{H}_B$ describes the interaction of $\hat{H}_0$ with the bath, and $\hat{H}_I$ describes the interaction of $\hat{H}_0$ with the applied electric field.\\

For TDBC molecules in a PVA host medium, $\hat{H}_B$ should represent the $3n_m-6=129$ intramolecular~\cite{Harris_Vibrational} vibrational modes (where $n_m$ is the number of atoms \textit{per} molecule) with a multitude of intermolecular modes. These vibrational modes are responsible for induced decay and dephasing in the system~\cite{McCumber_JAP_1963,Harris_JCP_1977}, along with a small shift in the excited state energy of the molecules~\cite{Ambrosek_JPChmA_116_2012}. Rather than determining $\hat{H}_B$ directly we have made a commonly-used simplification, that of incorporating the effects of the bath (vibrationally induced decay and dephasing) phenomenologically by application of the dissipative Lindblad superoperator (see below) and by making the assumption that the small energy shift can be ignored~\cite{Skinner_JPhysChem_1986,Abramavicius_Wurfel_2011,TKobayashiJAggregates2b}.\\

For an ensemble of $n$ two-level emitters (molecules), $\hat{H}_0$ can be written as~\cite{Valleau_JCP_137_034109_2012,Ambrosek_JPChmA_116_2012,TKobayashiJAggregates2},

\begin{equation}
	\hat{H}_0 = \hbar\omega_0|0\rangle\langle 0|+\sum_{i=1}^n\left (\hbar\omega^{(1)}_1|1_i\rangle\langle 1_i|+\mathop{\sum_{j=1}^n}_{j\neq i} J_{ij}|1_i\rangle\langle 1_j|\right ),
	\label{eq:Hamiltonian_elec}
\end{equation}

\noindent where $|0\rangle$ is the ground state of the nanoparticle, and $|1_i\rangle$ represents a single exciton excited in the nanoparticle, localized on molecule $i$, with the other molecules in their ground states \textit{i.e.} $|1_i\rangle=|0_1,...,1_i,...0_n\rangle$. In this way, only a single exciton is permitted within the ensemble at any time. The first term in brackets in \Eref{eq:Hamiltonian_elec}, $\hbar\omega_1^{(1)}$, represents the average energy eigenvalue of a non-interacting molecule in the excited state (an exciton). The second term corresponds to inter-molecular coupling, with coupling energy $J_{ij}$. The coupling is taken to be F\"{o}rster (dipole-dipole) coupling~\cite{YongShengZhaoOrganicNanophotonics,Valleau_JCP_137_034109_2012} since we assume that the overlap between the wave functions of each site are small. The corresponding interaction Hamiltonian $\hat{H}_I$ modeled in the Schr\"{o}dinger picture~\cite{Parker_1994} and written in the same basis is,

\begin{equation}
	\hat{H}_I=\sum^n_{i=1}\left (g_i^{*}|0\rangle\langle 1_i|+g_i|1_i\rangle\langle 0|\right ),	
	\label{eq:Hamiltonian_Int}
\end{equation}

\noindent where the coupling strength of the dipole to the external optical driving field is defined as $g_i=-\bi{E}(\bi{r}_i)\cdot\boldsymbol{\mu}_i$, where $\boldsymbol{\mu}_i$ is the exciton dipole moment.\\

Although thorough, a density matrix formed using Equations \eref{eq:Hamiltonian_elec}~\&~\eref{eq:Hamiltonian_Int} would have dimension $(n+1)\times (n+1)$, where $n$ is the number of molecules in the system. Given that $n$ can be several thousand for even a moderately-doped $100~nm$ diameter nanosphere, solving for such a large matrix would be computationally very demanding, despite considering only a single exciton in the ensemble. We therefore seek a simpler Hamiltonian for a TDBC-doped nanosphere which approximates the formalism above.\\

\noindent As a first step in this process, we identify that for an ensemble of aggregates (where the monomers within each aggregate are aligned with each other), the intra-aggregate coupling terms dominate~\cite{Knoester_2002}; this enables us to neglect the inter-aggregate coupling terms. By making this approximation, our approach to describe a nanoparticle doped with randomly distributed and randomly oriented aggregates is to first describe a Hamiltonian for a single aggregate, and then to take an orientational average.\\

\noindent The next step is to note that for a single aggregate, nearest-neighbour couplings dominate. The Hamiltonian matrix obtained under this approximation using \Eref{eq:Hamiltonian_elec} for a single aggregate containing $n$ monomers is,

\begin{equation}
H = \left (
	\begin{array}{cccccc}
		\hbar\omega_0 & 0 & 0 & 0 & \cdots & 0\\
		0 & \hbar\omega_1^{(1)} & J & 0 & \cdots & 0\\
		0 & J & \hbar\omega_1^{(1)} & J & \cdots & 0\\
		0 & 0 & J & \hbar\omega_1^{(1)} & \cdots & 0\\
		\vdots & \vdots & \vdots & \vdots & \ddots & \vdots\\
		0 & 0 & 0 & 0 & \cdots & \hbar\omega_1^{(1)}
	\end{array}\right ),
	\label{eq:Hamiltonian_elec_matrix}
\end{equation}

\noindent where $J$ is the nearest-neighbor interaction energy. The eigenvalues and eigenstates for this Hamiltonian matrix are derived in our Supporting Information. The first eigenstate is the ground state $|0\rangle$, with energy eigenvalue $\hbar\omega_0$. The second is a set of excited states where a single exciton is delocalized over the aggregate~\cite{Malyshev_PRB_51_1995,TKobayashiJAggregates2b}, 

\begin{equation}
	|m\rangle=\sqrt{\frac{2}{n+1}}\sum^n_{j=1}\sin\left(\frac{jm\pi}{n+1}\right)|1_j\rangle,
	\label{eq:excited_state_gen}
\end{equation}

\noindent where $1<m<n$. The single exciton transition dipole moment of the aggregate $\bi{d}_{01}(m)$ for mode $m$ is related to the transition dipole moment of the monomers $\boldsymbol{\mu}_{01}$ (assuming identical dipole moments) by~\cite{HochstrasserWhiteman_JCM_1972},

\begin{equation}
	\bi{d}_{01}(m) = \boldsymbol{\mu}_{01}\sqrt{\frac{1-(-1)^m}{n+1}}\cot\left (\frac{\pi m}{2(n+1)}\right ).
\end{equation}

\noindent This implies that $\bi{d}_{01}(m)$ is zero for even values of $m$. Even for very modest aggregates, $\sim n>6$, the leading eigenstate ($|1\rangle$) gives rise to a transition dipole moment a factor of three stronger than the next eigenstate, \textit{i.e.} the leading eigenstate is the 'brightest'~\cite{PhysRevA_53_2711}. We can take the transition dipole moment of the aggregate as $\bi{d}_{01}\approx\bi{d}_{01}(1)$, and use the two states $|0\rangle$ and $|1\rangle$ as an approximation for the aggregate, where $|1\rangle$, using equation 6, is given by,

\begin{equation}
	|1\rangle=\sqrt{\frac{2}{n+1}}\sum^n_{j=1}\sin\left(\frac{j\pi}{n+1}\right)|1_j\rangle.
	\label{eq:excited_state_gen2}
\end{equation}

\noindent The eigenvalue of $|1\rangle$ is (\textit{c.f.} Supporting Information, Equation S6),

\begin{equation}
	\hbar\omega_1 = \hbar\omega_1^{(1)}-2J\cos\left (\frac{\pi}{n+1}\right ).
\end{equation}

\noindent This allows us to write $\hbar\omega_1=\hbar\omega_1^{(1)}+\Delta$. The excitation energy of the aggregate is shifted from the monomer value by $\Delta$, and this shift arises from the interaction with other molecules in the aggregate. This energy shift has been observed elsewhere for aggregates~\cite{TKobayashiJAggregates,Ambrosek_JPChmA_116_2012}, and is loosely termed the `effect of aggregation'~\cite{YongShengZhaoOrganicNanophotonics}. The magnitude of $\Delta$ is typically hundreds of $meV$~\cite{Valleau_JCP_137_034109_2012}. Therefore, by considering only the ground state and this (brightest) excited state, \Eref{eq:Hamiltonian_elec} can be re-written as,

\begin{eqnarray}
	\hat{H}_0&\approx\hbar\omega_0|0\rangle\langle 0|+(\hbar\omega^{(1)}_1+\Delta)|1\rangle\langle 1|\nonumber\\
	 &=\hbar\omega_0|0\rangle\langle 0|+\hbar\omega_1|1\rangle\langle 1|.
	\label{eq:AggModEnergy}
\end{eqnarray}

\noindent The interaction Hamiltonian for the aggregate is written as,

\begin{equation}
	\hat{H}_I = G(|0\rangle\langle 1|+|1\rangle\langle 0|),
	\label{eq:potential}
\end{equation}

\noindent where $G=\bi{E}\cdot\bi{d}_{01}$ is the coupling strength of the electric field, $\bi{E}$, to the dipole moment, $\bi{d}_{01}$, of the aggregate. The Hamiltonian formed by adding \Eref{eq:AggModEnergy}~\&~\eref{eq:potential} can be applied to an ensemble of randomly-distributed aggregates by taking $G=\bi{E}\cdot\bi{\overline{d}}_{01}$, where $\bi{\overline{d}}_{01}=\bi{d}_{01}/\mathcal{D}$ is the orientational average of the aggregate dipole moments in the system of interest. Here, $\mathcal{D}$ is equal to either $2$ or $3$, corresponding to the number of spatial dimensions in the planar and bulk cases respectively (this orientational average is derived in our Supporting Information: see section 3).\\

Our goal now is to find an effective medium value of $\varepsilon$ at time $t$ and at the frequency of illumination, $\omega$. The first step is to note that $\hat{H}_I/|E(t)|$ defines the transition dipole matrix $\bi{d}$ for the system as a whole. Given that the density matrix ($\rho$) can be used to obtain the expectation value of an observable, we seek $\rho$ using the Liouville-von Neumann equation~\cite{Blum},

\begin{equation}
	\dot{\rho}(t,\omega) = -\frac{i}{\hbar}[H,\rho(t,\omega)]+L_D\rho(t,\omega).
	\label{eq:LvN}
\end{equation}

\noindent The first term in \Eref{eq:LvN} governs unitary evolution. The Lindblad dissipation superoperator~\cite{Schirmer_Solomon_PRA_70_022107_2004,Breuer_Petruccione_2002} $L_D$, is used to account for the decay and dephasing effects the bath has on the system. In this work, we assume the electron-phonon coupling to be weak at room temperature and for weak fields, and this enables the Born-Markov approximation upon which this formalism relies~\cite{PhysRevA.78.022106} to be used. The total dephasing rate of the transition $|0\rangle\leftrightarrow|1\rangle$ is $\Gamma_{01}$. This quantity is related to the population decay rate for the $|1\rangle\rightarrow|0\rangle$ decay channel, $\gamma_{01}$, and the pure dephasing rate, $\Gamma_{01}^{(d)}$, by~\cite{Schirmer_Solomon_PRA_70_022107_2004},

\begin{equation}
	\Gamma_{01} = \frac{\gamma_{01}}{2}+\Gamma_{01}^{(d)}.
	\label{eq:rates}
\end{equation}

\noindent The pure dephasing rate, $\Gamma_{01}^{(d)}$, arises from phase-changing interactions of the excitons with the environment~\cite{Wang_Chu_JChPhys_86_3225}, \textit{i.e.} the bath of vibrational modes. A less approximate approach could be adopted~\cite{PhysRevA.78.022106,Del_Pino_NJP_2015}, but assuming a simple rate for $\Gamma_{01}^{(d)}$ is sufficient for our present purposes, that of enabling an illustrative calculation to be carried out.\\

For our two-level system, \Eref{eq:LvN} is used to find $4$ coupled differential equations~\cite{Kavanaugh_Silbey}, the well-known Optical Bloch Equations~\cite{Allen_Eberly} (OBEs). Solving these for the applied cosine potential allows us to use the Rotating Wave Approximation (RWA)~\cite{prl_111_043601_Dorfman}, as detailed in our Supporting Information. To derive an expression for the permittivity, $\langle\bi{d}\rangle$ is determined using $\langle\bi{d}\rangle=\Tr(\rho\bi{d})$. For our aggregates, this value is equal to $\rho_{01}\overline{\bi{d}}_{01}+c.c.$. By choosing the forward-propagating electric field, \Eref{eq:Polarisation1} is re-arranged to give $\varepsilon$ for an ensemble of two-level molecules as,

\begin{equation}	
	\varepsilon(t,\omega)=\varepsilon_b+\frac{2N}{\varepsilon_0}\frac{|\bi{\overline{d}}_{01}|}{|\bi{E}|}\rho_{01}e^{i\omega t}.
	\label{eq:chi}
\end{equation}

\noindent This expression is applicable to an ensemble of molecules with number density $N$, arranged in aggregates, distributed randomly spatially and orientationally (in two or three dimensions), in a medium of background permittivity $\varepsilon=\varepsilon_b$ in the single-exciton regime. This formula holds for weak fields, as we show below. At first glance, \Eref{eq:chi} may appear to diverge for the case where $|\bi{E}|\rightarrow 0$. The resolution to this is that $\rho_{01}$ is linear in $|\bi{E}|$ in this limit: this gives us a field independent permittivity in this limit as expected.


\section{Results and Discussion}
\noindent For our model we require the following parameters: $\bi{\overline{d}}_{01}$, $\hbar\omega_1$, $\gamma_{01}$ and $\Gamma_{01}^{(d)}$. We used our experimental reflectivity and transmittance data (for a $1.46~wt\%$ TDBC:PVA $70~nm$ film~\cite{Gentile_NL_2014_14_2339}) to determine that $\hbar\omega_1=2.11eV\equiv 588~nm$. This agrees with the values obtained by van Burgel~\cite{van_Burgel_JChemPhys_1995_102_20} and Valleau~\cite{Valleau_JCP_137_034109_2012} although it is a slight change from our previous work, where we indicated that the transition occurred at $2.10~eV$ ($590~nm$), with a (weaker) shoulder transition at $2.03~eV$ ($610~nm$). Our revised value follows from an improved Kramers-Kronig analysis of our original data, as outlined in our Supporting Information.\\

From photoluminescence measurements~\cite{Wang_JPCL_5_14331439_2014}, we took the decay rate of $|1\rangle$ to be $\gamma_{01}=1.15\times 10^{12}s^{-1}$ for the aggregate in a PVA host medium. Using Molinspiration~\copyright, we determined the molecular weight and the effective volume of the TDBC molecule. Together with the concentration of the solution, these quantities allowed us to determine the molecular number density to be $N=1.47\times 10^{25}~m^{-3}$. We were then able to estimate the transition dipole moment for TDBC molecules in aggregate form, $\langle\bi{\overline{d}}_{01}\rangle$, and the dephasing rate, $\Gamma^{(d)}_{01}$, by fitting the steady-state solutions for \Eref{eq:chi} to our experimental data for $\varepsilon(\omega)$ by adjusting $|\bi{\overline{d}}_{01}|$ and $\Gamma^{(d)}_{01}$. In this way we found the dipole moment to be $48$ debye (D). The TDBC-doped thin films from which the experimental data were obtained were produced by spin-coating~\cite{Gentile_NL_2014_14_2339}. Previous work to investigate the orientation of dipole moments in thin polymer films produced by spin-coating found that the dipole moments lie predominantly in the plane~\cite{Garrett_Barnes_JMO_2004_51_2287}. Assuming that the TDBC aggregates also lie in the plane of the spun films reported in~\cite{Gentile_NL_2014_14_2339}, then the value of $48$ D we have determined here is a two-dimensionally averaged value, implying that the on-axis dipole moment of an aggregate is $\bi{d}_{01}=97$ D, and the three-dimensionally averaged moment is $32$ D. This three-dimensionally averaged moment compares with the $24$ D estimated by van Burgel \textit{et al.}~\cite{van_Burgel_JChemPhys_1995_102_20} from experiments in solution (3-dimensional).\\

\begin{figure}[!htbm]
	\includegraphics[width=\columnwidth]{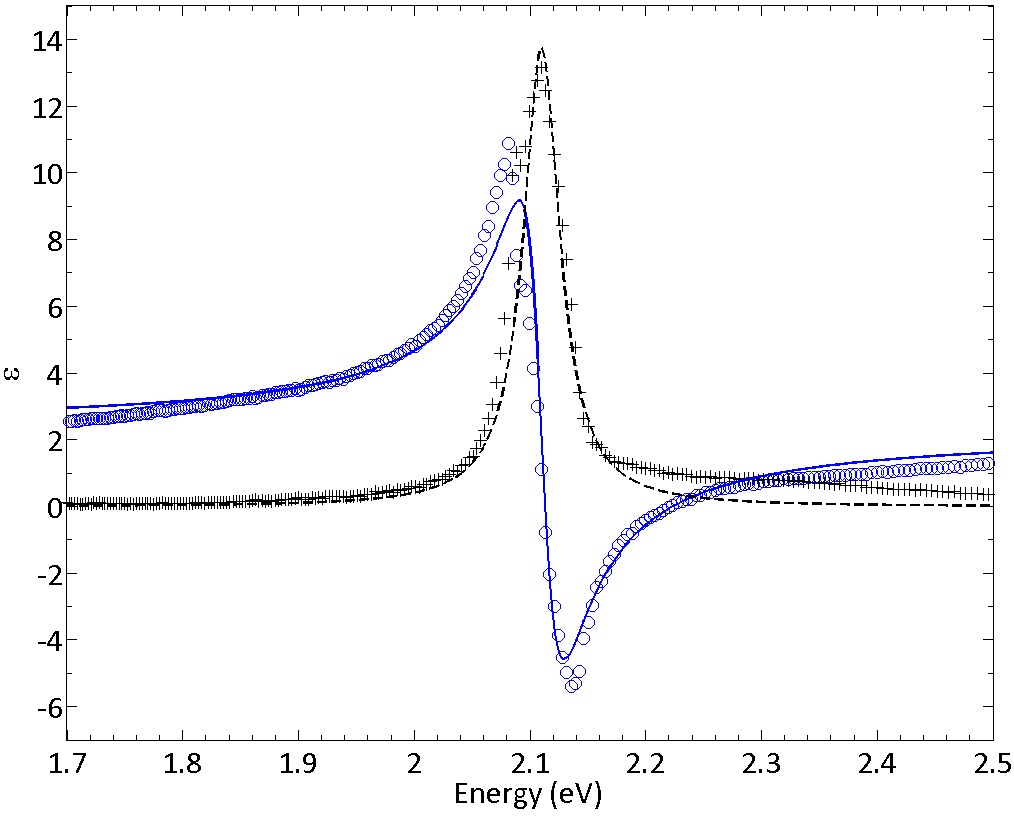}
	\caption{Extracted relative permittivity $\varepsilon(\omega)$ from a $1.46~wt\%$ TDBC:PVA film (circles for real part and crosses for imaginary part)~\cite{Gentile_NL_2014_14_2339}. The theoretical fit for $\varepsilon(\omega)$ indicated by the solid and dashed lines corresponds to a concentration of $1.46~wt\%$ ($3.22~wt\%$) for a two (three) dimensional distribution of dipoles.}
	\label{fig:2levelTDBCdielectric}
\end{figure}

\noindent The dephasing rate, $\Gamma^{(d)}_{01}$, was found to be equal to $17~meV$, which is $\approx\kappa_BT$ as expected~\cite{Valleau_JCP_137_034109_2012}. To provide additional support for our  value of $\Gamma^{(d)}_{01}$, we extracted and modeled $\varepsilon(\omega)$ from the reflectance and transmittance data from a $5.1~nm$ thick film obtained by Bradley \textit{et al.}~\cite{Bradley_AM_2005_17_1881}. We determined $\Gamma^{(d)}_{01}$ to be around $13~meV$, a value comparable with our own, bearing in mind that different bath spectral densities associated with differences in the host and substrate may change the value of $\Gamma^{(d)}_{01}$. Our results for $\varepsilon(\omega)$ against experimentally-determined data for our film are displayed in \Fref{fig:2levelTDBCdielectric}.

\noindent The permittivity of the thin film shown in \Fref{fig:2levelTDBCdielectric} has been modeled assuming the dipole moments of the aggregates are randomly oriented in the plane of the film. In what follows we wish to look at the optical response of a nanoparticle. For generality, and to ensure we consider an isotropic system, we will consider the dipole moments of the aggregates to be randomly oriented in three dimensions. Making this assumption requires us to increase the number density of our molecules from $1.47\times 10^{25}~m^{-3}$ to $3.29\times 10^{25}~m^{-3}$. In this way, our nanoparticle will be comprised of a material that has the same permittivity as that shown in \Fref{fig:2levelTDBCdielectric}. In all the calculations that follow we use this number density, which corresponds to a concentration of TDBC in PVA of $3.22~wt\%$.\\


\subsection{Numerical Results: Steady-State}\label{sec:SteadyStateResults}
We now explore theoretically the Mie~\cite{Mie,B+H} absorption efficiency spectra $Q_{abs}(\omega)$ for a $100~nm$ diameter nanosphere of $3.22\%$ TDBC:PVA, assuming a volume distribution of dipole orientations, based on $\varepsilon(\omega)$ calculated using \Eref{eq:chi}. In practice, the applied optical field we model here might be a laser beam. For a $1~mW$ laser with a spot diameter of $1.5~mm$, the strength of the electric field of our incident optical field would be equal to $462~Vm^{-1}$; we assume this value here.\\

In a $100~nm$ diameter nanosphere of our material, there are on average $n=1.72\times 10^4$ molecules. Note that it is the number of molecules and by extension their number density, which is the important quantity (and is used for $N$ in \Eref{eq:chi}) rather than the number density of aggregates, since each molecule provides a potential site for exciton excitation. To check the validity of our assumption that multi-exciton and nonlinear effects~\cite{Wang_Chu_JChPhys_86_3225} can be neglected, we computed the maximum expectation value of the number of excitons in the nanosphere ($n_{ex}=\max(\rho_{11})n$) using Equations \eref{eq:AggModEnergy} and \eref{eq:potential} in \Eref{eq:LvN}. We found that $n_{ex}/n\ll 1$ holds for laser powers of up to $10^2~W$ with a spot size of $1.5~mm$. Given that our laser power is $1~mW$, we assumed that the single-exciton linear regime is sufficient to describe the system under this illumination power.\\

\begin{figure}[!htbm]
\centering
	\includegraphics[width=\columnwidth]{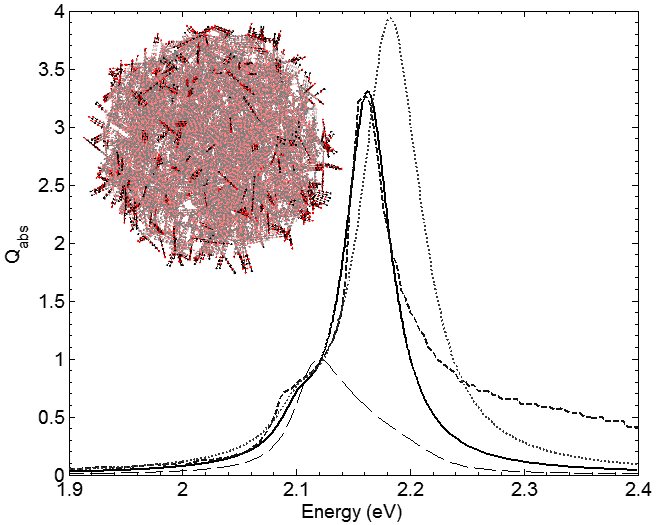}
	\caption{Mie calculations for the absorption efficiency $Q_{abs}(\omega)$ in the steady-state for a $100~nm$ diameter nanosphere of $3.22~wt\%$ TDBC:PVA using the values for $\varepsilon(\omega)$ from experiment (dashed line), from two-level OBEs (solid line), and from our previous Lorentz oscillator model (dotted line). The material absorption coefficient $\kappa$ (imaginary part of the refractive index), normalized to unity, is also plotted  for illustrative purposes (long dashed line). Inset: a 3D representation of the aggregated emitters (assuming brick-stone aggregation, with 15 molecules \textit{per} aggregate) randomly distributed in a $100~nm$ diameter nanosphere.}
	\label{fig:figure2}
\end{figure}

\noindent In \Fref{fig:figure2} we plot the absorption efficiency $Q_{abs}(\omega)$ for a $100~nm$ diameter nanosphere, calculated for a variety of permittivities; in each case the absorption efficiency is calculated using Mie theory \cite{Mie,B+H}. Calculated values for $Q_{abs}$ based upon the permittivity obtained using our improved analysis of experimental data are shown in \Fref{fig:figure2} as a dashed line. Our quantum theoretical spectrum for $Q_{abs}$, using $\varepsilon(\omega)$ from \Eref{eq:chi}, is shown as the solid line. This theoretically derived spectrum provides a close match to the extracted data, most importantly for energies in the region of interest below $2.22~eV$. For energies exceeding $2.2~eV$, there is a limb in the extracted data (dashed curve) which might perhaps be attributed to inhomogeneous (non-Lorentzian) broadening which is not accounted for using the OBEs. Also displayed in \Fref{fig:figure2} is the result for $Q_{abs}$ using a best-fit classical Lorentz oscillator model (the parameters for which can be found in our Supporting Information) shown as a dotted line. It can be seen that the quantum model outlined in the present paper provides an improved fit to the experimental data. We attribute this to the inclusion of dephasing ($\Gamma_{01}^{(d)}$) in the model: if $\Gamma_{01}^{(d)}$ were set to zero, a Lorentz model would be recovered in the steady state, and the single damping term in the Lorentz model would have to accommodate both decay and dephasing. Therefore, by including dephasing, the actual physical value of the decay rate $\gamma_{01}$ can be included to achieve an accurate result for $\varepsilon$.\\

\begin{figure}
	\includegraphics[width=\columnwidth]{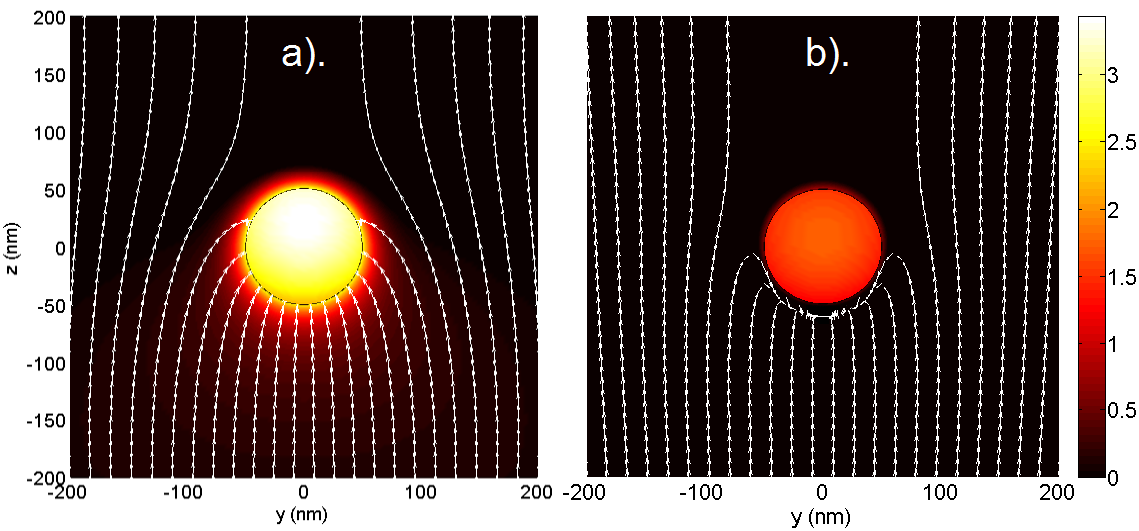}
	\caption{The time-averaged electric field strength normalized to the incident field strength (color plot) and the Poynting vector $\bi{S}$ (arrows) in the vicinity and on the surface of the $100~nm$ $3.22~wt\%$ TDBC:PVA nanosphere, with incident power flow along the positive z-direction. Data are calculated for incident photon energies of a). $2.16~eV=574~nm$ and b). $2.12~eV=586~nm$ corresponding to peak absorption efficiency and $\kappa$ respectively.}
	\label{fig:figure3a_3b}
\end{figure}

It is interesting to note a key feature shown by the data in \Fref{fig:figure2}: $Q_{abs}$, reaches its peak value at $2.16~eV$ ($574~nm$). This is in contrast to the absorption coefficient, $\kappa(\omega)$, which peaks at $2.12~eV$ ($586~nm$), shown as a long dashed line in \Fref{fig:figure2}. This difference in spectral position arises because the peak in $Q_{abs}$ is not due simply to absorption: rather, it is due to the excitation of a localized SEP mode~\cite{Gentile_NL_2014_14_2339}. Confirmation of this interpretation comes from two sources. First, in the quasistatic limit the polarizability of the nanosphere follows the Clausius-Mossotti condition, for which resonance occurs when $\varepsilon$ is real-valued and equal to $-2$ (when the nanosphere is in free space)~\cite{NandH}. From \Fref{fig:2levelTDBCdielectric} this can be seen to be approximately true for our absorbing $100~nm$ diameter nanosphere, as the permittivity value at the wavelength of peak absorption efficiency, $Q_{abs}$, is complex and equal to $\varepsilon=-2.251+1.728i$. This difference from $\varepsilon=-2$ originates from the fact that $\varepsilon=-2$ only gives the resonance condition if the imaginary part of $\varepsilon$ is zero; the complex nature of the permittivity changes the spectral location of the absorption peak. Second, $Q_{abs}$ near the peak goes well above unity: this is associated with field enhancement~\cite{Vollmer}, another signature of a resonant mode. The enhanced electric field in the vicinity of the nanosphere is illustrated graphically in \Fref{fig:figure3a_3b}(a), together with direction of power flow shown by the Poynting vector $\bi{S}$.\\

\noindent In ~\Fref{fig:figure3a_3b}, the incident electric field is polarized in the x-direction. The Poynting vector arrows shown in the figure were calculated at starting points for which $z=-200~nm$ and $x=0~nm$, linearly spaced in the range $-200~nm\leq y\leq 200~nm$. Subsequent points for evaluation of the Poynting vector were taken at $10~nm$ steps in the direction of the Poynting vector at each point, resulting in the flux lines shown. The power flow in \Fref{fig:figure3a_3b}(a) shows that incident light is drawn towards and absorbed by the nanosphere for starting positions up to around $130~nm$ from the central position of the nanosphere. This demonstrates that at this energy, the nanosphere absorbs more light than the light geometrically striking it~\cite{BohrenAmJPhys_1983}, and hence $Q_{abs}>1$. In comparison, absorption at the transition energy, \textit{i.e.} at $2.11~eV$ is seen only as a shoulder mode in the absorption efficiency of the nanosphere (\Fref{fig:figure2}) and the efficiency does not exceed unity. The power flow around the nanosphere for the energy at which $\kappa$ peaks ($2.12~eV$) is shown in \Fref{fig:figure3a_3b}(b), and the enhancement of the field is much weaker than for excitation on resonance at $2.16~eV$.

\subsection{Numerical Results: Time Domain}\label{sec:TDBCtd}
We now turn our attention to the time domain. Our theoretical model for dynamic processes in two-level quantum systems subject to a perturbing cosine electric field is similar to models considered elsewhere~\cite{Fox,Foot,Slowik_PRB_88_195414_2013,OptComm_283_23532355_2010}, but here the observable of interest arises from the temporal evolution of the coherences of the density matrix, rather than the populations. The dynamics of a two-level ensemble subject to a pulse potential has been the subject of recent investigation~\cite{Sukharev_ACSNANO_8_807817_2014}, but here we investigate a rather different case: that of a cosine potential of fixed amplitude that is switched instantaneously on at some moment in time. We do this to provide an easily soluble model that illustrates the time-dependent phenomena we wish to discuss.\\

\begin{figure}[!htbm]
\centering
	\includegraphics[width=\columnwidth]{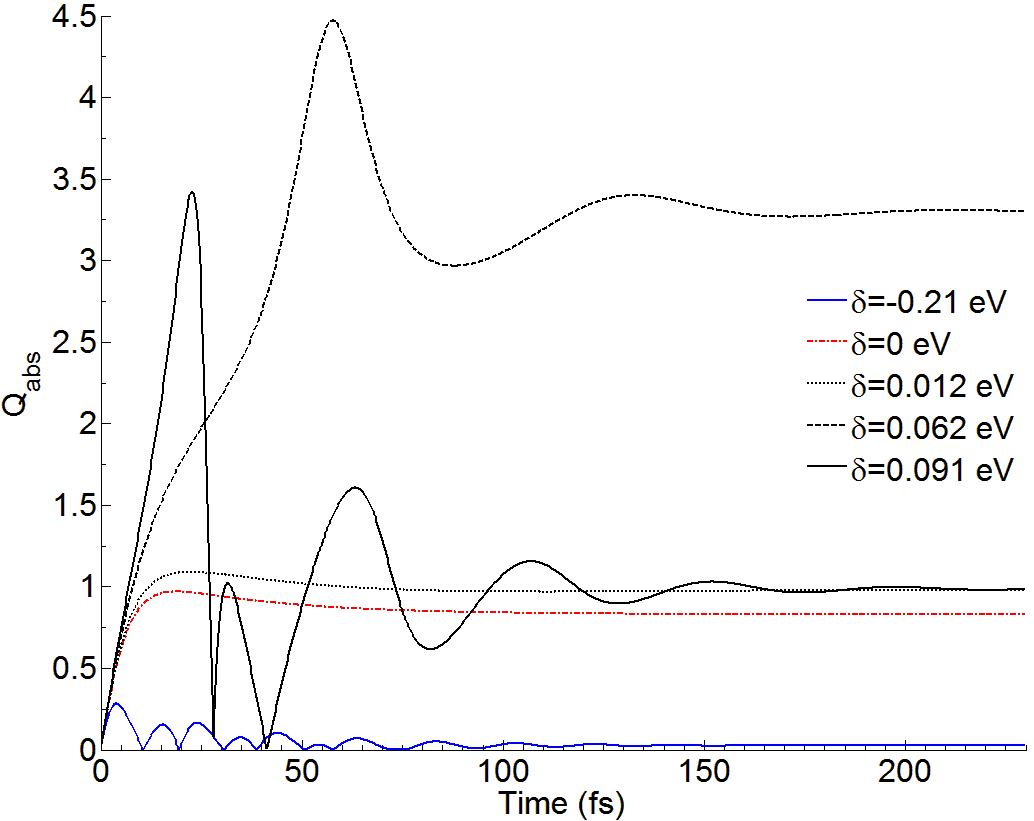}
	\caption{Calculated $Q_{abs}(t,\omega)$ for a $100~nm$ diameter nanosphere of $3.22\%$ TDBC:PVA warming up a pure state at $t=0$ with a $1~mW$ laser set at five different detunings $\delta$ from the exciton transition.}
	\label{fig:figure4}
\end{figure}

By using \Eref{eq:chi} to calculate $\varepsilon(t)$ for a given illumination frequency $\omega$ as before, $Q_{abs}(t,\omega)$ can be determined and its temporal behavior examined. To do this, we again use Mie theory. This is an approximation since the fields scattered in Mie theory are assumed to be instantaneous. Given that the dynamics seen in \Fref{fig:figure4} evolve over a few femtoseconds and that light propagates over a length scale three times the size of the nanoparticle during a single femtosecond, and that the nanoparticle is illuminated with an electric field of constant amplitude, this approximation is deemed to hold. Mie theory can therefore be used to give a quasi-instantaneous picture of the absorption.\\

$Q_{abs}(t)$ is shown in \Fref{fig:figure4} for five different detunings, $\delta=\hbar\omega_1-\hbar\omega$, from the transition at $\hbar\omega_1=2.11~eV$. We assume that all the molecules in the nanoparticle are initially in their ground state. At $t=0$ we turn on our field abruptly. We see that a steady-state response is attained after  $>200~fs$, but interestingly, for $\delta=0.09~eV$, $Q_{abs}(t)$ repeatedly exceeds unity in spite of the steady-state value of $Q_{abs}$ being below unity at this detuning. Since $Q_{abs}(t)>1$ implies field enhancement, these data are indicative of a transient LSEP mode being present at early times. The time-dependent behavior comprises two contributions: the first is the oscillatory behavior arising from Rabi oscillations; the second is the transient effects associated with the sudden turning-on of the field. In the latter, the magnitude of the density matrix coherences exceed their steady-state values for tens of femtoseconds, resulting in larger values of $\varepsilon$ and hence different absorption properties to the steady-state.\\

If $\varepsilon$ passes through the Clausius-Mossotti condition for the nanoparticle as it approaches its steady-state value, $Q_{abs}$ exceeds unity, implying a transient LSEP mode. This is seen best for a detuning of $0.091~eV$ between $0-30~fs$. The Rabi oscillations follow the generalized Rabi frequency $\tilde{\Omega}_R$, given by~\cite{PhysRevLett_46_1192},

\begin{equation}
	\tilde{\Omega}_R=\sqrt{\Omega_1^2+\delta^2},
	\label{eq:generalRabi}
\end{equation}

\noindent where, $\Omega_1\ll\delta$, and $\tilde{\Omega}_R\rightarrow\delta$ in this case, $\Omega_1$ and $\delta$ are the Rabi frequency and the detuning respectively. These Rabi oscillations are naturally convoluted with the transient effects. This implies, together with the short timescales involved in the system, that it would be a challenge to see these transitory effects, but might perhaps be possible~\cite{Vasa_NatPhot_2013_7_128}. Critical to the transient LSEP lifetime is $\Gamma^{(d)}_{01}$. If this dephasing could be reduced without losing the transient negative permittivity that is essential for field enhancement (and field confinement), then the transient timescale of the system would be increased up to a maximum of $1/\gamma_{01}$. This corresponds to the picosecond regime for our TDBC:PVA system. Under this circumstance, transient LSEP modes would become more easily observable.\\

\section{Conclusions}\label{sec:conclusions}
We have re-evaluated the measurements reported in our previous work and have obtained an improved permittivity for our J-aggregate-doped $1.46~wt\%$ TDBC:PVA polymer film. Using a quantum-mechanical framework we have given support to our previous investigation based on a classical analysis~\cite{Gentile_NL_2014_14_2339}, that TDBC doped nanoparticles can exhibit a localized surface exciton-polariton (LSEP) mode. We have used a quantum model to show that these nanoparticles may also exhibit transient LSEP modes in the sub-picosecond regime. These results help strengthen the idea that molecular excitonic materials provide an interesting alternative upon which to base nanophotonics~\cite{Saikin_Nanophotonics_2013_2_21}. By using molecular materials the possibility of bottom-up approaches such as supramolecular chemistry and self-assembly can be brought to bear on the production of nanophotonic structures.


\ack{The work was supported in part by the UK Engineering and Physical Sciences Research Council, and in part by The Leverhulme Trust.}

\section*{References}
\bibliography{2Level_TDBC}

\end{document}


\title[Localized exciton-polariton modes in dye-doped nanospheres]{Localized exciton-polariton modes in dye-doped nanospheres: a quantum approach\\ Supporting Information}

\author{Martin J Gentile, Simon A R Horsley, William L Barnes}
\address{School of Physics and Astronomy, University of Exeter, Exeter EX4~4QL, UK}
\ead{\mailto{m.j.gentile@exeter.ac.uk}, \mailto{s.horsley@exeter.ac.uk}, \mailto{w.l.barnes@exeter.ac.uk}}

\section{Derivation of the excited states of an aggregate}
The eigenvalues of the matrix Hamiltonian in our main text (Equation 5) are solutions to the equation,

\begin{equation}
\left |
	\begin{array}{cccccc}
		\hbar\omega_0-\lambda & 0 & 0 & 0 & \cdots & 0\\
		0 & \hbar\omega_1^{(1)}-\lambda & J & 0 & \cdots & 0\\
		0 & J & \hbar\omega_1^{(1)}-\lambda & J & \cdots & 0\\
		0 & 0 & J & \hbar\omega_1^{(1)}-\lambda & \cdots & 0\\
		\vdots & \vdots & \vdots & \vdots & \ddots & \vdots\\
		0 & 0 & 0 & 0 & \cdots & \hbar\omega_1^{(1)}-\lambda
	\end{array}\right |=0,
	\label{eq:Hamiltonian_matrix}
\end{equation}

\noindent where $J$ represents the nearest-neighbour coupling. \Eref{eq:Hamiltonian_matrix} can be expressed in the form,

\begin{equation}
\left |
	\begin{array}{cc}
		\hbar\omega_0-\lambda & 0\\
		0 & H_n\\		
	\end{array}\right |=0,
\end{equation}

\noindent where $H_n$ is the sub-matrix written explicitly in \Eref{eq:Hamiltonian_matrix}, and is representative of an aggregate with $n$ molecular units. The eigenvalues of \Eref{eq:Hamiltonian_matrix} can be determined by solving,

\begin{equation}
	(\hbar\omega_0-\lambda)|H_n|=0.
\end{equation}

\noindent The first of these is the ground state energy $\lambda_0=\hbar\omega_0$, which readily yields the eigenvector $|0\rangle$. The other eigenvalues and eigenvectors require more consideration. The first step is to find an expression for $|H_n|$. Writing this determinant in terms of further sub-matrices gives the following recursive relationship,

\begin{equation}
	|H_n| = (\hbar\omega_1^{(1)}-\lambda_m)|H_{n-1}|-J^2|H_{n-2}|.
	\label{eq:recursive1}
\end{equation}

\noindent The next step is to identify that this recursive relationship can be put into the same form as the recurrence relation for Chebyshev polynomials~\cite{Chebyshev_1854} \textit{i.e.}

\begin{equation}
	U_n(x)=2xU_{n-1}(x)-U_{n-2}(x).
	\label{eq:Chebychev_rec}
\end{equation}

\noindent In doing this, and after a little algebra, an expression for the eigenvalues $\lambda_m$ is determined,

\begin{equation}
	\lambda_m = \hbar\omega_1^{(1)}-2J\cos\left (\frac{m\pi}{n+1}\right ),
\end{equation}

\noindent where $1<m<n$. The first of these ($\lambda_1$) is the excited state of the aggregate ($\hbar\omega_1$) taken in our main text.\\

The eigenvectors $|m\rangle$ are determined by analysis of $H_n|m\rangle$. The $j^{th}$ element of $H_n|m\rangle$ is,

\begin{equation}
	Jm_{j-1}+\hbar\omega_1^{(1)}m_j+Jm_{j+1}=\lambda_m m_j,
\end{equation}

\noindent which can also be put into the form of \Eref{eq:Chebychev_rec}. Making this identification and a little algebra yields the following expression for the normalised excited states of the aggregate~\cite{Malyshev_PRB_51_1995,TKobayashiJAggregates2b},

\begin{equation}
	|m\rangle = \sqrt{\frac{2}{n+1}}\sum_{j=1}^n\sin\left (\frac{jm\pi}{n+1}\right )|1_j\rangle.
\end{equation}

\section{Solving the Optical Bloch Equations}
The Optical Bloch Equations (OBEs) derived from the Liouville von-Neumann equation can be written in the compact form as,

\begin{equation}
	\dot{\vec{\rho}} = \bar{L}\vec{\rho},
\end{equation}

\noindent where $\vec{\rho}$ is a vector of the density matrix elements. One may solve the OBEs by application of the Rotating Wave Approximation (RWA)~\cite{prl_111_043601_Dorfman} with subsequent use of a matrix inversion method, or numerically by way of a Runge-Kutta method, such as the RK10(8) method~\cite{Hairer}. In this section, both of these approaches are detailed, evaluated, and shown to be equivalent for an ensemble illuminated with a cosine potential.

\subsection{Matrix inversion method}
The matrix-inversion method relies on application of the rotating wave approximation (RWA)~\cite{arXiv13013585} to the perturbing potential, which enables one to write $\bar{L}(t,\omega)$ as a time-independent matrix, $\bar{L}(\omega)$. In this case, solutions for the density matrix elements can be written in the following form,

\begin{equation}
	\rho_{mn}(t,\omega) = \sum_i c_{mn,i}e^{i\omega_{i}t},
	\label{eq:rho_super}
\end{equation}

\noindent using the principle of superposition. The coefficients $c_i$ are determined from initial conditions and from the eigenvectors of $L(\omega)$. The angular frequencies $\omega_i$ are related to the eigenvalues $\lambda_i$ of the matrix $L(\omega)$ by $i\omega_i=\lambda_i$. $\lambda_i$ is complex with a negative real part in order to conserve probability. The RWA may be applied to a cosine potential, $G= \boldsymbol{d}\cdot\boldsymbol{E}_0~cos(\omega t)$. An advantage of this method is that computation time is very short and weak fields may be considered easily.\\

A restriction on the RWA is that only a field of one frequency may impinge upon the system in this method, since the time-independence of $\bar{L}$ must be preserved. In order to solve the OBEs for circumstances including a frequency-spread pulse, a numerical approach must be used instead, as we now indicate.

\subsection{Explicit Runge-Kutta methods}
For the general case where $\bar{L}$ cannot be written as time-independent \textit{e.g.} when the system is subjected to a pulse, a numerical method must be used to solve the OBEs. The OBEs have the general form $\dot{y}=f(t,y)$ which has a general solution written in discretised form,

\begin{equation}
	y_{n+1} = y_n + h_n\sum^n_{i=1}{b_i k_i},\\
	\label{eq:RK_solution}
\end{equation}

\noindent where,

\begin{equation}
	k_i = f\left (t_n+h_na_i,y_n+h_n\sum^{j=i-1}_{j=1}{C_{ij}k_j} \right ).
\end{equation}

\noindent The estimated error at each step is evaluated as,

\begin{equation}
	e_{n+1} \propto y_{n+1} - y^*_{n+1}.
\label{eq:RK45_error}
\end{equation}

\noindent The values $a_i$, $b_i$ and $C_{ij}$ all depend upon the specific method involved. The original 1st-order algorithm using this general method is the Euler method~\cite{Euler}, but a much more accurate method is the Runge-Kutta (RK4) method, which has been used previously to probe the dynamics of 2-level systems~\cite{Charron_JChemPhys_138_024108_2013}. This method suffers from being non-adaptive, in the sense that the step size is always taken to be constant. This makes numerical solutions for rapidly-changing behaviour unreliable. An improved approach is to dynamically allocate the step size between each iteration through a comparison of the estimated local error between the 4th and 5th-order solutions at each point. The yields the Runge-Kutta-Fehlberg (RK4(5)) method~\cite{Fehlberg}. The advantage of an adaptive numerical method such as the RK4(5) method over a non-adaptive method such as the RK4 method is that rapidly-changing behaviour can be modelled with greater accuracy. Local errors are also minimized and the resultant solution that one determines is numerically smoother~\cite{Tong_JCompAppMath_233_10561062_2009}. However, for our work the RK4(5) method is insufficient for producing numerically stable solutions for weak fields. The RK10(8) method (a 10th-order Runge-Kutta method with in-built error estimation by comparison to the 8th order)~\cite{Hairer} was tested and found to produce smoother output for insignificantly more computing time than that of the RK4(5) method.\\

The main advantage of a Runge-Kutta method over the RWA is that no terms in the Hamiltonian are neglected. As a result, any arbitrary potential can be modelled, not just a cosine potential. However, the main drawback of any Runge-Kutta method is that the computation time for the process is many times longer than using the RWA and matrix inversion. This longer timescale is sometimes prohibitive depending on the parameters involved. A further drawback is that for weak fields the solutions to the OBEs become stiff in time and numerical instability is encountered. This can be seen from the fact that with our calculations for TDBC illuminated with a $1~mW$ cosine

\begin{figure}
	\centering
	\includegraphics[width=\columnwidth]{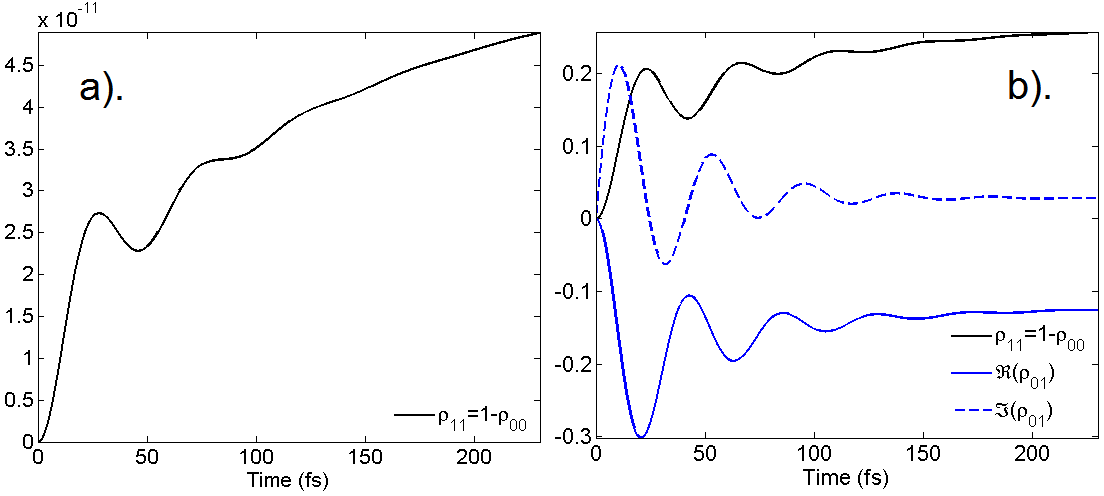}
	\caption{a). The relative occupancy of the excited state $\rho_{11}$, for a $1~mW$ laser potential with spot size $1.5~mm$. b). The relative occupancy of the excited state $\rho_{11}$, together with the coherence $\rho_{01}$ (real and imaginary parts shown) for a $10~MW$ laser with the same spot size.}
	\label{fig:figurea1a_a1b}
\end{figure}

\noindent potential with spot size $1.5~mm$, the relative occupancy of the excited state $\rho_{11}$ does not exceed one part in $10^{10}$ and so $\rho_{00}$ remains more-or-less constant in time with value $\approx 1$ as shown in \Fref{fig:figurea1a_a1b} (conversely, the coherences oscillate in time at the applied frequency and are of order $10^{-6}$). Therefore, the populations are incredibly stiff and this is where explicit numerical methods fail. One way to avoid the resultant numerical instability is to confine this procedure to solving OBEs with strong fields. To achieve physically meaningful solutions in this circumstance, one would need to introduce other effects such as multi-exciton recombination and nonlinear effects. By neglecting these effects, similar solutions can be obtained to those of weak fields for the coherences, provided the field does not induce significant population inversion. As an alternative to this compromise, one can use an implicit Runge-Kutta method to solve stiff equations. However, the computation time for implicit Runge-Kutta methods was found to exceed that of the explicit methods by at least a factor of ten, making this approach ungainly.\\

In summary, our procedure is as follows: for weak fields the RWA is used. Solutions derived using this method are supported by the solutions obtained using the RK10(8) method for strong fields with multi-exciton and nonlinear effects neglected. We conclude that by careful use of the RWA, realistic behaviour can be simulated subject to the condition that only one frequency illuminates the system. For slowly time-varying strong fields or for two strong lasers, an explicit Runge-Kutta method may be used, but for weak time-varying fields an implicit Runge-Kutta method may be implemented successfully.
%
%
\section{The permittivity of a collection of randomly oriented non--interacting dipoles}
The quantum mechanical model discussed in the main text predicts the polarizability of an individual aggregate.  A typical macroscopic sample consists of a large number of randomly oriented aggregates that can be treated as non-interacting.  The permittivity of such a macroscopic sample has a simple relationship to the microscopic aggregate polarizability that depends on the dimensionality of the sample, which we derive in this section.\\

For a collection of $N$ dipoles with dipole moments $\bi{d}_{i}$ the polarization per unit volume $\bi{P}$ is given by,

\begin{equation}
	\bi{P}=\frac{1}{V}\sum_{i=1}^{N}\bi{d}_{i}=\frac{N}{V}\left\langle\boldsymbol{\alpha}(\omega)\right\rangle\cdot\bi{E}=\boldsymbol{\chi}(\omega)\cdot\bi{E}.\label{polarization}
\end{equation}

\noindent In the second step of the above equation we assumed that the dipole moments are induced by an electric field that can be treated as uniform over the volume $V$, and that the dipoles have a tensor polarizability $\boldsymbol{\alpha}_{i}(\omega)$ that depends on the frequency of the field.  The average value of the polarizability is defined as $\langle\boldsymbol{\alpha}(\omega)\rangle=N^{-1}\sum_{i}\boldsymbol{\alpha}_{i}(\omega)$.  If we assume that the dipoles are of the same type then the polarizabilities $\boldsymbol{\alpha}_{i}(\omega)$ have the same magnitude $\alpha(\omega)$, and differ only due to the dipole orientation $\hat{\boldsymbol{n}}_{i}$.  For a collection of such dipoles distributed in a $\mathcal{D}$ dimensional space,

\begin{eqnarray}
	\langle\boldsymbol{\alpha}(\omega)\rangle&=\alpha(\omega)\langle\hat{\bi{n}}\otimes\hat{\bi{n}}\rangle\nonumber\\[10pt]
	&=\alpha(\omega)\left(\langle n_{x}^{2}\rangle\hat{\bi{x}}\otimes\hat{\bi{x}}+\langle n_{y}^{2}\rangle\hat{\bi{y}}\otimes\hat{\bi{y}}+\dots\right)\nonumber\\[10pt]
	&=\frac{\alpha(\omega)}{\mathcal{D}}\left(\hat{\bi{x}}\otimes\hat{\bi{x}}+\hat{\bi{y}}\otimes\hat{\bi{y}}+\dots\right)\label{average_alpha}
\end{eqnarray}

\noindent where to obtain the second and third lines we used isotropy to determine that quantities such as $\langle n_{x}n_{y}\rangle$ are zero and that $\langle n_{x}^{2}\rangle=\langle n_{y}^{2}\rangle=\dots$, and used the normalization of the unit vectors \(\hat{\bi{n}_{i}}\cdot\hat{\bi{n}}_{i}=1\) to determine that \(\langle n_{x}^{2}\rangle+\langle n_{y}^{2}\rangle+\dots=\mathcal{D}\langle n_{x}^{2}\rangle=1\).  Combining Equations \eref{polarization} and \eref{average_alpha} we can then find the macroscopic permittivity $\varepsilon$ which is isotropic and equal to,

\begin{equation}
	\varepsilon(\omega)=1+\frac{N\alpha(\omega)}{\mathcal{D}}
\end{equation}

\noindent with $\mathcal{D}=2$ for a planar sample, and $\mathcal{D}=3$ for a bulk one.
%
%

\section{Improved Analysis of Experimental Data}
\begin{figure}[!htbm]
	\centering	
		\includegraphics[width=\columnwidth]{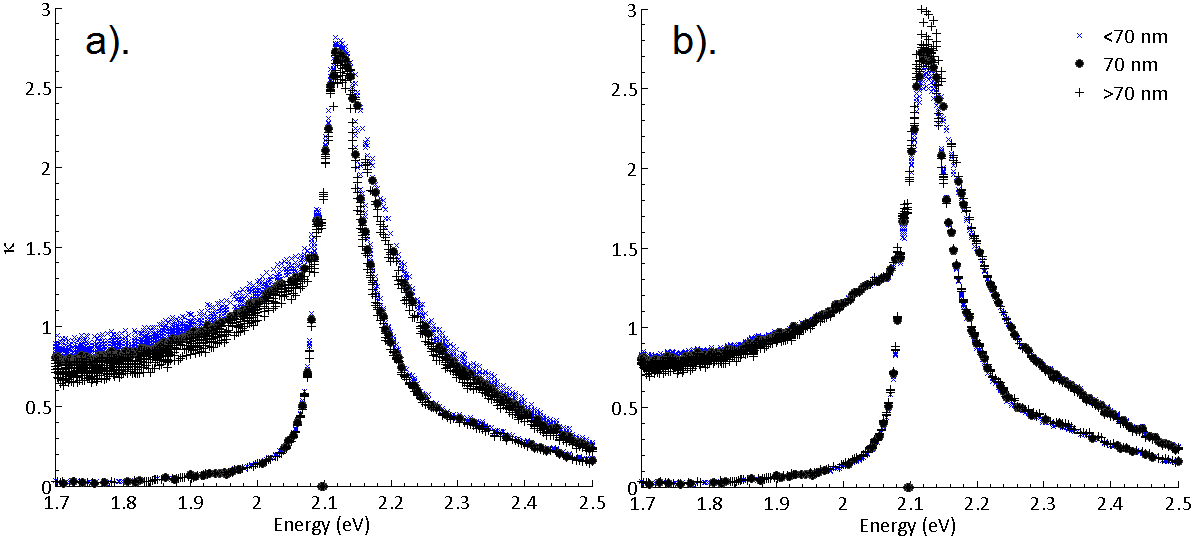}				
	\caption{$\kappa$ for a $1.46 wt\%$ TDBC:PVA film, assuming a range of thicknesses before (a) and after (b) adjustment for these thicknesses.}
	\label{fig:figurea2a_a2b}
\end{figure}

To extract the complex relative permittivity $\varepsilon$ or equivalently, the complex refractive index $\tilde{n}=n+i\kappa$, of our thin film, we compared our experimental values of reflectance ($R_e$) and transmittance ($T_e$) at normal incidence with theoretical ones for each wavelength, following the theoretical framework outlined by Heavens~\cite{Heavens}. In this formalism, the function,

\begin{equation}
	f(n,\kappa)=|T_t(n,\kappa)-T_{e}|+|R_t(n,\kappa)-R_{e}|,
	\label{eq:fresnelresidual}
\end{equation}

\noindent must be minimised, where the subscript $t$ denotes theoretical values. This process relies upon perfect values of $R_e$ and $T_e$ and $n$ and $\kappa$ are `guessed' to lie in a sensible range. Eq.~\ref{eq:fresnelresidual} is minimised over this range. By this process, small errors in $R_e$ and $T_e$ coupled with rounding errors in the guessed range of values for $n$ and $\kappa$ can yield spurious final values of $n$ and $\kappa$. In order to improve this process, we made additional calculations for $\kappa$, varying the thickness of the film from $63-77 nm$, the range of experimental uncertainty. Our results are shown collectively in \Fref{fig:figurea2a_a2b}(a).\\

\noindent In this figure, we have kept the first two minimum values of Eq.~\ref{eq:fresnelresidual} for each wavelength. Therefore, the figure shows the physical and first spurious solutions. The gaps associated with taking a single thickness are evidence for the need to consider other thicknesses. Doing this produces a range of values for $\kappa$ for each wavelength. We then adjusted each value for the thickness taken by assuming for a given thickness $t$,

\begin{equation}
	T\approx e^{-\kappa t}=e^{-\kappa't'},
\end{equation}

\noindent which leads to,

\begin{equation}
	\kappa'=\frac{t}{t'}\kappa.
\end{equation}

\begin{figure}[!htbm]
	\centering	
		\includegraphics[width=0.5\columnwidth]{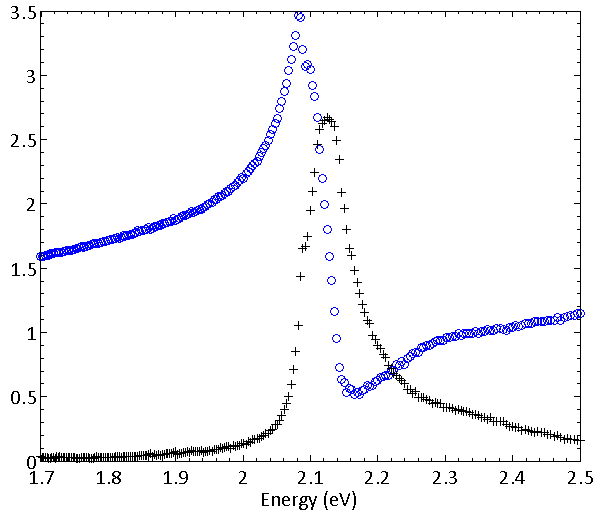}				
	\caption{The improved extracted values of the real (blue circle) and imaginary (black pluses) parts of the refractive index from experimental values.}
	\label{fig:figurea3}
\end{figure}

\noindent This has the effect of deconvolution on our data, to produce \Fref{fig:figurea2a_a2b}(b). The two zero-values show where the process has still failed. Our final step was to eliminate the spurious values for $\kappa$, and use the Kramers-Kronig relations to find $n$, as shown in \Fref{fig:figurea3}.\\

\noindent This process produces much smoother values of $\tilde{n}$ than in our previous work. This is because we assumed previously that the true value of $\kappa$ lay in the middle of the two solutions found in \Fref{fig:figurea2a_a2b} and the input thickness was held constant. With this improved analysis by varying the input thickness, it can be seen that intermediate values between the two clear solutions in \Fref{fig:figurea2a_a2b} do not exist, and it becomes apparent that the upper solution is a spurious one.\\

\newpage
A single Lorentz oscillator model can be used to describe the permittivity of a material classically~\cite{Fox2,lebedev},

\begin{equation}
	\varepsilon(\omega) = \varepsilon_m + \frac{f_0\omega_0^2}{\omega_0^2-\omega^2-i\omega\gamma_0}.
\end{equation}

\begin{figure}
	\includegraphics[width=\columnwidth]{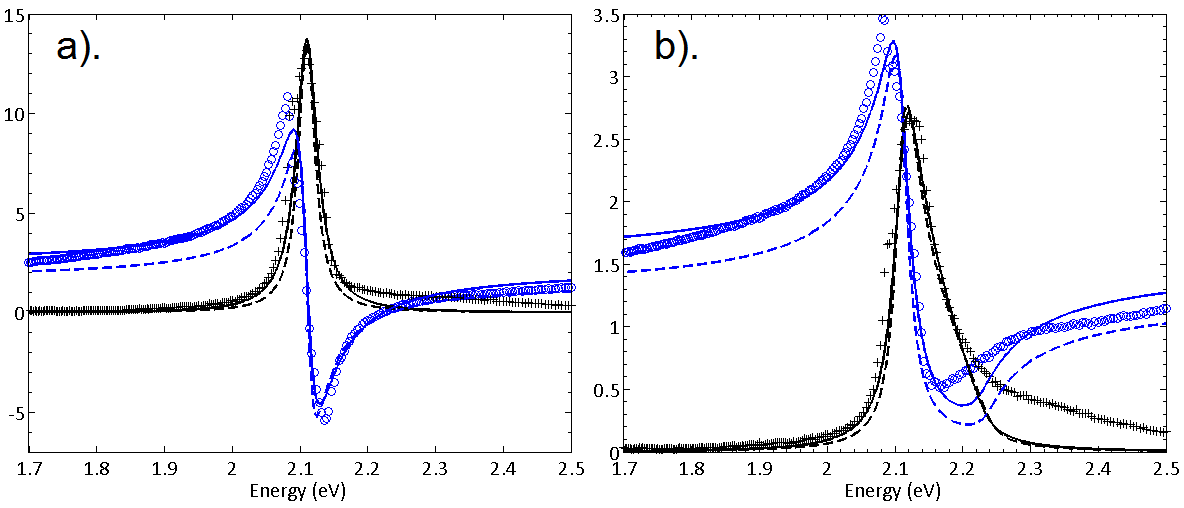}
	\caption{The real (blue) and imaginary (black) parts of the a). permittivity and b). complex refractive index of our $1.46~wt\%$ TDBC:PVA film. Solid lines correspond to results from Optical Bloch Equations and dashed lines correspond to results from a single-oscillator Lorentz model. The discrete data points correspond to our extracted values.}
	\label{fig:figurea4a_a4b}
\end{figure}

\noindent The parameters for this model which fit our data in Fig.~\ref{fig:figurea3} best (assuming a host medium of $n_m=1.52$) are $\omega_0=2.11~eV$, $f_0=0.3$, $\gamma_0=46.1~meV$. The output of these for $\varepsilon$ and the complex refractive index $\tilde{n}$ are illustrated in \Fref{fig:figurea4a_a4b} together with our quantum fit and extracted data. It can be seen that the Lorentz model is not as close to the extracted data as the quantum model, particularly for the real parts of both $\varepsilon$ and $\tilde{n}$.

\section*{References}
\bibliography{2Level_TDBC}